IAC-12-E3.2.8

# THE UNITED NATIONS HUMAN SPACE TECHNOLOGY INITIATIVE (HSTI) ACTIVITY STATUS IN 2012

**Mika Ochiai**
United Nations Office for Outer Space Affairs, Austria, mika.ochiai@unoosa.org

**Aimin Niu, Hans Haubold, Takao Doi**
United Nations Office for Outer Space Affairs, Austria
aimin.niu@unoosa.org  hans.haubold@unoosa.org  takao.doi@unoosa.org

In 2010, the Human Space Technology Initiative (HSTI) was launched by the United Nations Office for Outer Space Affairs (UNOOSA) within the United Nations Programme on Space Applications. The Initiative aims at promoting international cooperation in human spaceflight and space exploration-related activities, creating awareness among countries on the benefits of utilizing human space technology and its applications, and building capacity in microgravity education and research. HSTI has conducted a series of outreach activities and expert meetings bringing together participants from around the world. HSTI will also be implementing science and educational activities in relevant areas to raise the capacities, particularly in developing countries, in pursuit of the development goals of the United Nations, thus contributing to promoting the peaceful uses of outer space.

## I. INTRODUCTION

More than 50 years have passed since Yuri Gagarin achieved the first human spaceflight on 12 April 1961, opening a new chapter of human endeavour in outer space. The years that followed saw the rapid development of space science and technology in the realm of human presence in space, including the historic first Moon landing by Neil Armstrong on 20 July 1969, and the development of several space stations such as the Soviet Union's Salyut and Mir and the U.S.'s Skylab programme on low Earth orbit. The first Apollo-Soyuz docking project was a significant step for furthering the collaborative human endeavour in space.

In 1998, the construction of the International Space Station (ISS) began with the combined efforts of 15 nations [1]. With its size of 110 meters long, 74 meters wide, weighing almost 420 metric tons, the ISS offers unique and cooperative opportunities in both scientific and engineering projects. Humanity has maintained the multinational permanent human presence in outer space aboard the ISS, gaining new knowledge, technological innovation and creating invaluable assets for future generations.

### I.I United Nations Committee on the Peaceful Uses of Outer Space

Back in 1959, the United Nations General Assembly established the Committee on the Peaceful Uses of Outer Space (COPUOS) in order to review the scope of international cooperation and to devise programmes in this field, undertaken under the United Nations' auspices, as well as to encourage continued research and dissemination of information on outer space matters, and to study legal problems arising from the exploration of outer space [2].

With its two subsidiary bodies - the Scientific and Technical Subcommittee and the Legal Subcommittee, COPUOS deals with a wide range of issues concerning current and future activities in space, and reports to the General Assembly which adopts an annual resolution on international cooperation in the peaceful uses of outer space [3]. Several multilateral treaties have been adopted by the General Assembly to enable the orderly conduct of activities in outer space. The cornerstone of these governance instruments is the Treaty on Principles Governing the Activities of States in the Exploration and Use of Outer Space, including the Moon and Other Celestial Bodies, known as "Outer Space Treaty" of 1967 [4]. One of the other four treaties, the "Rescue Agreement" of 1968, requires States to assist an astronaut in case of accident, distress, emergency or unintended landing [5].

In 2011, the Declaration on the 50[th] Anniversary of Human Space Flight and of COPUOS, contained in the General Assembly resolution 55/2 [6], stressed the need to look more closely into how advanced space research and exploration systems and technologies could further contribute to meeting challenges, and endeavour to examine how the outcomes and spin-offs of scientific research in human space flight could increase the benefits, in particular for developing countries.

I.II United Nations Programme on Space Applications

At the first United Nations Conference on the Exploration and Peaceful Uses of Outer Space (UNISPACE '68) [7], held in Vienna in 1968, Member States recommended the creation of a dedicated programme in the framework of the United Nations to support countries which lacked the human and technical resources necessary to fully utilize the benefits of space technology. In its resolution 2601 A (XXIV) of 16 December 1969, the General Assembly endorsed "the recommendation of the Committee on the Peaceful Uses of Outer Space for the appointment by the Secretary-General of a qualified individual with the full-time task of promoting the practical applications of space technology" [8]. The initial activities - in what was later to be called the "United Nations Programme on Space Applications" - commenced in May 1971. The United Nations Office for Outer Space Affairs (UNOOSA), as the Secretariat for COPUOS, was also given the responsibility for implementing the Programme.

Following the second UNISPACE conference held in 1982 (UNISPACE '82), the mandate of the Programme was broadened to further develop capacities for research and applications of space science and technology [9].

The third and the latest UNISPACE Conference, held in 1999 (UNISPACE III) [10][11], touched upon, for the first time, the significance of the International Space Station (ISS) for humanity and the importance of encouraging international cooperation, particularly cooperation and partnerships between countries operating and already using the ISS and developing countries [12].

Since its inception in 1971, the United Nations Programme on Space Applications has provided support to countries in making full use of the benefits of space science and technology and their applications for social and economic development. The overall strategy of the Programme is to focus on selected areas that are critical, particularly for developing countries, defining and working towards objectives achievable in two to five years and build on the results of previous activities [13]. From 1971 to 2010, the Programme conducted over 270 activities such as expert meetings, seminars, workshops, and international conferences in 67 different countries with more than 18,000 participants [14]. The current priority areas are: (a) environmental monitoring, (b) natural resources management, (c) satellite communications for tele-education and telemedicine applications, (d) disaster risk reduction, (e) developing capabilities in the use of global navigation satellite systems (GNSS), (f) basic space science, including the International Space Weather Initiative, (g) space law, (h) climate change, (i) the Basic Space Technology Initiative, and (j) the Human Space Technology Initiative [15].

## II. HUMAN SPACE TECHNOLOGY INITIATIVE (HSTI)

II.I Objectives

The Human Space Technology Initiative (HSTI) was launched by UNOOSA in 2010 [16] as the latest initiative under the framework of the United Nations Programme on Space Applications. HSTI was built on the relevant recommendations [17] related to human spaceflight and exploration contained in the report of UNISPACE III:

*400. International partnerships and cooperation between countries and companies involved in the operation and utilization of the International Space Station and those countries not yet participating in that endeavour should be encouraged;*

*401. Information about utilization of the International Space Station should be disseminated throughout the world in order to increase awareness of the matter in countries not yet participating in that endeavour;*

*402. Mechanisms for improving accessibility from a technical and financial point of view (for example, loans from the World Bank) should be encouraged to simplify utilization of the International Space Station, especially for developing countries."*

Taking into particular account the latest progress in the world as well as the needs of developing countries for building capacity and providing opportunities in this area, the objectives of HSTI were defined as follows:
- To promote international cooperation in human spaceflight and space exploration-related activities;
- To create awareness among Member States on the benefits of utilizing human space technology and its applications; and
- To build capacity in microgravity education and research.

II.II Methodology and implementation plans

A work plan spanning three years from 2011-2013 was set up, encompassing the following activities: (1) to provide a forum for exchanging information on human space technology and its applications, (2) to inform Member States about utilization opportunities of the ISS and other facilities; and (3) to support Member States in increasing their capacity in microgravity research and education.

The first action comprises annual expert meetings, workshops, and symposiums in different regions of the world. This will bring experts together in order to establish a common view on how to facilitate HSTI. These meetings will be complemented by periodic outreach activities with the purpose of providing information in the relevant fields as well as HSTI's activities to a broader audience.

The second action involves the publication and distribution of informative materials to institutions and schools. This aims at increasing awareness of the possibilities and benefits of microgravity-related science and technology development.

The last action includes science activities and the distribution of appropriate educational materials to students, teachers and researchers with the purpose of increasing their capacities in relevant fields. The most substantial part of this action will be the distribution of zero-gravity instruments to selected universities and high-schools, particularly in developing countries, in order to provide them with a low-cost and easy way to conduct research under conditions similar to microgravity.

Figure 1 shows the timeline of the work plan as currently updated.

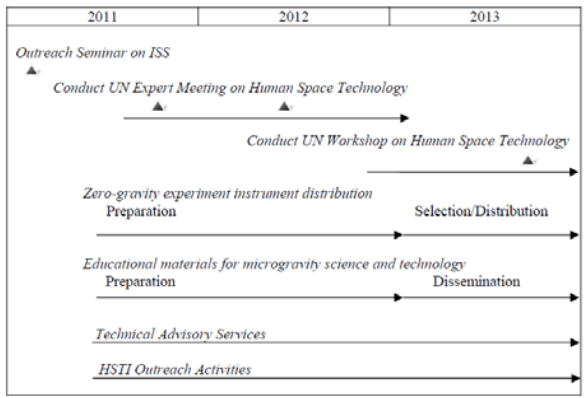

Fig. 1: Human Space Technology Initiative Work Plan (2011-2013)

### III. HSTI ACTIVITY STATUS

Based on the work plan in Figure 1, an outreach seminar focusing on ISS's research and educational activities, and an expert meeting with over a hundred participants were respectively organized in 2011. These led to a follow-up expert meeting in June 2012 aimed at extending the benefits of ISS utilization.

III-I Outreach Seminar on the International Space Station on 8 February 2011

As the first activity of HSTI, UNOOSA organized a one-day activity called "Outreach Seminar on the International Space Station (ISS)" in Vienna on 8 February 2011, during the 48th Session of the Scientific and Technical Subcommittee of COPUOS [18]. With the purpose of disseminating information on the activities carried out onboard the ISS to a broader community, as well as facilitating discussions on HSTI, the seminar brought together a total of 17 countries including developing nations. The participants shared the status of various activities and accomplishments aboard the ISS, and information on cooperation opportunities between ISS partners and interested institutions.

A mechanism called the Non-Partner Participation policy, developed by the ISS partnership in 2002, governs how non-ISS partners can participate in ISS research. A non-partner should first form a team with one of the five ISS partners (Canadian Space Agency (CSA), European Space Agency (ESA), Japan Aerospace Exploration Agency (JAXA), National Aeronautics and Space Administration (NASA), and Russian Federal Space Agency (Roscosmos)) and the ISS partnership would then review the bilateral cooperation for approval. Non-partners are encouraged to contact one of the ISS partners with their collaborative ideas [19].

After exchanging views among participants, the seminar concluded that HSTI could be a meaningful mechanism for raising awareness on the potential of ISS research and educational activities among countries, regions, and potential users that have up to this point not been involved in such activities, thereby contributing to capacity building in microgravity science and technology education .

III-II Expert Meeting on Human Space Technology on 14-18 November 2011

The United Nations/Malaysia Expert Meeting on Human Space Technology was held in Malaysia from 14 to 18 November 2011. This was the first United Nations event of its kind to discuss how the world can benefit from human space technology, and to further develop international collaborative activities in human space exploration. Hosted by the Institute of Space Science of the National University of Malaysia, co-organized by UNOOSA and the five ISS partner agencies, more than 120 experts from both developing and industrialized nations participated in the meeting [20].

Throughout the five-day meeting, various presentations on the following topics were delivered: ISS programmes; microgravity science; education, outreach and capacity-building; and national, regional, and international space programmes. These were followed by discussions in three working group sessions: Microgravity Research; Education, Outreach and Capacity-Building; and HSTI. Participants provided remarks and observations on those themes with the final objective being to develop shared recommendations on the Initiative.

On-orbit human facilities such as the ISS can and have been providing an ideal microgravity environment for research and experiments to better understand fundamental scientific questions and to provide solutions for problems on Earth, in the fields of physics as well as fluid, material, life, and engineering science. The utilization of ground-based research facilities, such as clinostats, drop towers, parabolic flights and centrifuges can, however, facilitate microgravity research and is essential as a preparatory step towards in-flight experiments.

As a tool for capturing and cultivating interest in science and sparking imagination among students, "space education" would be presented in a more attractive way and in a language adapted to the target group. Owing to ISS's long-term operation, a considerable number of students have been and would be reached by educational projects to utilize the ISS. Projects such as "Seeds in Space" and "Amateur Radio on the ISS" are cited as examples. Over 400,000 students have studied golden orb spiders living on ISS. Additionally, astronauts have frequently supported educational activities, for example, by giving talks in schools or carrying out educational demonstrations.

A presentation on ISS benefits for humanity reported that the ISS Partnership had recently identified the following three areas in which activities on ISS could benefit humanity – education, Earth observation and disaster response, and human health. To facilitate cooperation in extending benefits of ISS research and education to the world, potential partnership activities between ISS and HSTI were also brought up for further consideration.

China has also been progressively carrying out their manned space missions since 1990s. The potential cooperation within the framework of HSTI, composed of three offers, was presented by the China Manned Space Engineering Office. These included flight experiment opportunities on board Chinese space laboratories for scientists and engineers from around the world as well as an international astronaut programme including astronaut selection, training, and flight opportunities. Cooperation in capacity-building was also proposed in the areas of human space technology and its applications.

At the end of the meeting, the following 10 recommendations were endorsed by the participants [21]:

(a) The Human Space Technology Initiative should take action to create awareness among stakeholders, including decision makers in the public and private sectors, researchers and students, of the social and economic potential of space science and technologies and to initiate outreach activities;

(b) The Initiative should identify and inform Member States about space-related research opportunities and organize meetings in which invited experts can provide information to interested parties;

(c) The Initiative should establish capacity-building programmes, including through the provision of educational material, instrument distribution and/or access, national or regional expert centres, training of trainers, exchange programmes and competition and motivation programmes;

(d) The Initiative should serve as a catalyst for international collaboration by promoting the formation of common interest groups, conducting regular surveys of countries concerning their space competence profiles, developing a set of guidelines for collaboration, promoting multinational institutional agreements and creating regional expert centres;

(e) The Initiative should promote the exchange of knowledge and the sharing of data by raising awareness, promoting user-friendly mechanisms for data access and providing knowledge about self-supporting habitats for application, including for energy efficiency on Earth;

(f) Governments, institutions and individuals are encouraged to use space-based platforms for research in the following areas: psychology and social interaction in a multicultural, confined and isolated environment; vaccine development; nutritional, agricultural and food security; human physiology and aging; space technology for future exploration; and the space environment;

(g) Governments, institutions and individuals are encouraged to explore ground-based research for gravity-related science, preparing space-based experiments and making use of microgravity simulators (such as clinostats), microgravity instruments (such as parabolic flights, drop tubes and drop towers), hyper-gravity instruments (such as centrifuges) and software models;

(h) Governments, institutions and individuals are encouraged to explore the opportunities for commercial alternatives for educational and research activities in space, such as sub-orbital flights and long-duration experiments;

(i) Governments and institutions are encouraged to use space education as a tool for inspiring and motivating people and sustaining interest in science and technology;

(j) Governments are encouraged to incorporate space education into the curricula of schools (in different subjects such as mathematics, physics, biology, chemistry and social science) and universities.

III-III Expert Meeting on the International Space Station Humanitarian Benefits on 11-12 June 2012

From 11-12 June 2012, the United Nations Expert Meeting on ISS humanitarian benefits was held in Vienna, Austria, during the 49th Session of COPUOS [22]. With the aim of extending its benefit to all in the following identified areas - Earth observation and disaster response, health, and education [23], the meeting brought together experts from ISS partners and United Nations agencies to discuss and identify potential collaborations.

The agencies participated from ISS partners were NASA, CSA, ESA and JAXA, and the specialized UN agencies were the World Meteorological Organization (WMO), the United Nations Environment Programme (UNEP), the World Health Organization (WHO), and the United Nations Educational, Scientific and Cultural Organization (UNESCO).

Throughout the two-day meeting, activities onboard the ISS and those pursued by the UN were presented to serve as the grounds for identifying the potential ways of extending the benefits.

*Earth observations and disaster response:* ISS offers a unique vantage point for observing the Earth's ecosystems, which cover about 85% of the Earth surface and 95% of the Earth's population [24]. A variety of Earth-observation payloads have been and will be attached to the ISS. The "Crew Earth Observations (CEO)" has collected imagery for various ground, coastal, and atmospheric targets including dynamic events and disasters, in support of collaborative science, education, public outreach and disaster response. The Solar Monitoring Observatory onboard the ISS provides detailed measurements of the Sun's spectral irradiance and can contribute to climate modelling of the Earth's environment. Other studies include bird migration tracking where a multinational project for tracking small animals on a global scale uses miniaturised tags that communicate with ISS receivers and transmitters.

The WMO Space Programme has among its main objectives to enhance the capabilities and interoperability of the global space-based observing system. One of the ideas raised was a radiometric calibration reference payload to be installed onboard the ISS, which could provide an ISS-based calibration reference standards that would enhance the quality of information for wide range Earth observation. UNEP's central tasks encompass global environment monitoring, bringing issues and potential solutions to the attention of governments and international communities for action. UNEP has been analysing the state of global environment, providing early warning information, and assessing environmental trends at both regional and global levels. UNESCO has been supporting its World Heritage sites in utilizing satellite observation technology in cooperation with space agencies. UNESCO's global network of universities could also help extend ISS benefits to more people and more countries.

*Health:* The ISS and earthbound analogues provide unique possibilities to study the reaction of the human body under extreme environmental conditions. Studies have been conducted in a microgravity environment on balance disorders, cardio vascular de-conditioning, decrease of bone mineralization and muscle disuse atrophy. Other research areas include studies of a confined and multi-cultural environment, the effects of cosmic radiation linked to cancer risk, and the reduction of immune responses. Medical systems developed for ISS crew members were also referred to as potentially contributing to improved healthcare on Earth. A training and remote-guidance tool for portable ultrasound device, developed by NASA in cooperation with universities and hospitals, allows non-physician astronauts to rapidly diagnose and treat a wide variety of medical conditions. This tool, already modified for terrestrial applications, can be used to benefit vulnerable populations on a larger scale. The technology of robotic arms for the ISS has led to the world's first MRI (Magnetic Resonance Imaging) compatible image-guided, computer-assisted device specifically designed for neurosurgery. Other examples of benefits include: telemedicine advancements, macromolecular crystallization, and water recycling technology.

WHO has the expertise and experience to provide criteria to identify telemedicine applications for underserved populations on Earth. In line with the WHO's Global Observatory for eHealth and the call for innovative health technologies, one of the collaborative areas discussed was the identification of ISS-proven technologies and services and its transfer that are suitable for low-resource settings.

*Education:* A considerable number of students have benefitted from educational projects that utilize the ISS. "Butterflies, Spiders and Plants in Space", conducted by NASA from 2008-2012, has demonstrated the effectiveness of using the ISS as a platform for student-centred experiments and education in science, technology, engineering and mathematics learning. During the experiments, organism life cycles and behaviour in microgravity were recorded in still images and video which were available online globally along with teaching guides. The ISS also offers a variety of educational information. Modular didactic materials, produced by ESA, include ISS education kits available in 12 languages and movie materials covering basic space science, health and nutrition education, and space robotics. Online lessons for both primary and secondary students in 13 languages as well as courses for university students and professors are also available. Space Poem Chain, led by JAXA, connects people, including crew members in space, by allowing them to think together about the universe, Earth, and life itself, and create a linked verse.

One potential area, proposed by UNOOSA, is the distribution of educational material, translated into six UN official languages for science, technology, engineering and mathematics. Taking advantage of the United Nations' expertise and global network of institutions and schools, several ideas were brought up in such areas as the distribution of user-friendly educational contents, student-led projects where ISS would be an educational platform.

In order to convert these collaborative ideas to possible future projects, further assessment by the interested parties will be needed. As a result, collaborative projects may be initiated in those areas.

III. IV Zero-Gravity Instrument Distribution Project

Based on the recommendations of the UN Expert Meeting on Human Space Technology in Malaysia in 2011, HSTI will be carrying out science activities aimed at enhancing capacity-building in the areas of microgravity science, particularly in developing countries [25].

The "Zero-gravity Instrument Distribution Project", the primary science activity of HSTI, is now underway, in which one-axis clinostats will be distributed worldwide to selected institutions of higher education. This project is expected to provide unique opportunities for students and researchers in observing the growth of indigenous plants in their countries in a simulated microgravity condition. The target groups for distribution are high-schools, universities and laboratories, particularly in developing countries. A one-axis clinostat, along with a teacher's guide, was selected for distribution on the basis of its ease of use as well as the potential scientific benefits of the project.

One cycle is scheduled for two years from the announcement of opportunity to the submission of the final activity report. Currently, two cycles are being scheduled, distributing 15 clinostats per cycle. The experiment phase will last approximately a year, during which institutions will use the clinostat to conduct experiments on the proposed projects using indigenous plant species. One of the conditions of the selection will be to provide annual reports on their activities with the clinostat to UNOOSA. The project is furthermore expected to create datasets of plant species with their gravity response, which would be used to design future space experiments as well as contribute to the advancement of science.

In order to select suitable institutions to receive the clinostats and increase the scientific value of the project, the HSTI Science Advisory Board (HSTI-SAB) will be established, comprised of several experts in microgravity life science. HSTI-SAB's primary task is to advise HSTI in the selection of institutions based on the evaluation of their educational or research proposals.

Based on the scientific evaluation of the final reports submitted at the end of the 1st and 2nd cycles and the status of support from Member States, HSTI may extend the project to its 3rd and 4th cycles.

IV. CONCLUSION

In 2010, the Human Space Technology Initiative (HSTI) was launched in the framework of the United Nations Programme on Space Applications, for the purpose of raising awareness of the benefits of human space technology, promoting international cooperation in activities related to human space flight and space exploration, and building capacity in microgravity research and education, particularly in developing countries.

In 2011, the outreach seminar on the International Space Station and the UN Expert Meeting on Human Space Technology were organized both of which led to the basis of future HSTI activities.

In 2012, the UN Expert Meeting on the ISS Humanitarian Benefits, as the first of its kind in the United Nations, took place involving relevant UN agencies to extend the benefits of the ISS for humanity. In order to increase the capacity to conduct microgravity-related research, capacity building activities centred on distribution of the zero-gravity instruments are soon to be carried out, particularly in developing countries.

Over the last 50 years of space exploration, human space technology has become an essential part of the advancement of civilization. HSTI strives to bring the benefits of human space activities to all and to bring nations together for this endeavour, thus, creating new opportunities for international cooperation.

ACKNOWLEDGEMENT AND DISCLAIMER

We are grateful to all who joined the activities of HSTI, especially to those who took part in the meetings and seminars in 2011 and 2012, and who provided constructive comments to proceed with the HSTI implementation.

The views expressed herein are those of the authors and do not necessarily reflect the views of the United Nations.